\documentclass[11pt,a4paper,english]{article}
\pdfoutput=1

\usepackage{lmodern}
\renewcommand{\sfdefault}{lmss}

\usepackage{float}
\usepackage{mathtools}
\usepackage{amsmath}
\usepackage{amsthm}
\usepackage{amssymb}
\usepackage{stmaryrd}
\usepackage{enumitem}

\newif\ifColors
\newif\ifShowSigns

\ifx\ifCrop\undefined\else
  \Colorstrue
  \ShowSignstrue
\fi

\makeatletter

\newif\ifprstyle

\ifx\affiliation\undefined
  \prstylefalse
\else
  \prstyletrue
  \newif\ifnotoc
  \notoctrue
\fi

\ifprstyle
  \usepackage[
    colorlinks=true, pdfa=true,
    urlcolor=blue,   anchorcolor=blue,
    citecolor=blue,  filecolor=blue,
    linkcolor=blue,  menucolor=blue,
    pagecolor=blue,  linktocpage=true
  ]{hyperref}
\else
  \usepackage{jheppub}
  \newcommand{\email}[1]{\emailAdd{#1}}
\fi

\ifx\ifCrop\undefined\else

  \ifx\ifPresentation\undefined
     \usepackage[paperwidth=171.2mm,paperheight=261.2mm]{geometry}
  \else
     \usepackage[paperwidth=330mm,paperheight=233mm]{geometry}
  \fi

  \divide\@tempdima\baselineskip
  \@tempcnta=\@tempdima
  \ifx\ifPresentation\undefined
     \hypersetup{pdfstartview={Fit},pdfpagelayout={TwoPageRight}}
  \else
    \hypersetup{pdfstartview={Fit},pdfpagelayout={SinglePage}}

    \usepackage{everypage}
    \AddEverypageHook{\tikz[remember picture,overlay]{%
      \draw [gray!10] ($(current page.east)+(-130mm,+180mm)$)
         -- ($(current page.east)+(-130mm,-100mm)$);}}
  \fi
\fi

\hypersetup{
    urlcolor=black!40!blue,   anchorcolor=black!40!blue,
    citecolor=black!40!blue,  filecolor=black!40!blue,
    linkcolor=black!40!blue,  menucolor=black!40!blue,
    pagecolor=black!40!blue
}

\usepackage{bookmark}

\usepackage{amsmath,amsfonts,amssymb,bm,cancel}

\usepackage{color,xcolor}
\usepackage{colortbl}

\usepackage{tensind}
\tensordelimiter{?}

\ifprstyle
  \newcommand{\repEqq}[2]{\tag{\ref{#1}#2}}
\else
  \newcommand{\repEqq}[2]{\tag{\ref*{#1}#2}}
\fi

\newcommand{\bSe}{\begin{subequations}}
\newcommand{\eSe}{\end{subequations}}

\newcommand{\bWe}{\begin{widetext}}
\newcommand{\eWe}{\end{widetext}}

\newtheorem{proposition}{Proposition}

\allowdisplaybreaks

\usepackage{tikz}
\usetikzlibrary{shapes,arrows}
\usetikzlibrary{calc}
\usetikzlibrary{positioning}
\usetikzlibrary{decorations.pathreplacing}
\usetikzlibrary{shapes.misc}

\usepackage[scr,scaled=1.1]{rsfso}
\usepackage{eucal}

\DeclareMathAlphabet{\mathsfit}{\encodingdefault}{\sfdefault}{m}{sl}
\SetMathAlphabet{\mathsfit}{bold}{\encodingdefault}{\sfdefault}{bx}{sl}

\usepackage{float}

\floatstyle{plaintop}
\newfloat{bbox}{thp}{lop}
\floatname{bbox}{Box}

\renewcommand\floatc@plain[2]{\setbox\@tempboxa\hbox{{\@fs@cfont #1.} #2}%
\ifdim\wd\@tempboxa>\hsize {\@fs@cfont #1.} #2\par
\else\hbox to\hsize{\hfil\box\@tempboxa\hfil}\fi}

\ifprstyle\else
  \usepackage[toc]{multitoc}
  
  \addtocontents{toc}{\protect\small}
\fi

\newcommand{\emitFrontMatter}{
  \ifprstyle
    \begin{abstract}\myAbstract\end{abstract}
  \else
    \ifnotoc
      \abstract{\\[-1.64ex]\hphantom{\hspace{0.28cm}}%
        \parbox{1\columnwidth-0.28cm}%
        {\renewcommand\baselinestretch{1.05}\small\hspace{18mm}\myAbstract}}
    \else
      \abstract{\myAbstract}
    \fi
  \fi

  \keywords{\myKeywords}

  \ifprstyle\else
    \makeatletter
      \def\@fpheader{\myReleaseInfo}
    \makeatother
    \ifnotoc
      \compress
      \renewcommand\afterLogoSpace{}
      \renewcommand\afterSubheaderSpace{}
      \renewcommand\afterProceedingsSpace{}
      \makeatletter
      \renewcommand\ps@titlepage{}
      \makeatother
      \toccontinuoustrue
    \else
      \renewcommand\afterTocSpace{\medskip}
      \addtocontents{toc}{\protect\setcounter{tocdepth}{2}}
    \fi
  \fi

  \maketitle

  \ifprstyle\else
    \ifnotoc
      \hrule\bigskip\bigskip
    \else
      \flushbottom
    \fi
  \fi
}

\usepackage{titlesec}

\newcommand{\emitAppendix}{
  \phantomsection
  \addcontentsline{toc}{section}{Appendices}
  \addtocontents{toc}{\protect\setcounter{tocdepth}{2}}

  \makeatletter
    \def\toclevel@section{1}
    \def\toclevel@subsection{2}
  \makeatother

  \addtocontents{toc}{\protect\makeatletter}
  \addtocontents{toc}{\string\let\string\l@chapter\string\l@section}
  \addtocontents{toc}{\string\let\string\l@section\string\l@subsection}
  \addtocontents{toc}{\protect\makeatother}

  \titleformat{\section}{\normalfont\large\bfseries}{Appendix~\thesection~~~}{0em}{}

  \let\oldSection\section
  \renewcommand\section{\bookmarksetupnext{level=2}\oldSection}

  \appendix
}
\makeatother

\usepackage{babel}
\begin{document}

\ifx \ii \undefined  


\definecolor{red}{rgb}{1,0,0.1}
\definecolor{green}{rgb}{0.0,0.6,0}
\definecolor{blue}{rgb}{0.1,0.1,1}
\definecolor{orange}{rgb}{0.6,0.3,0}
\definecolor{magenta}{rgb}{0.9,0.1,1}

\newcommand\nPlusOne{$N$+1}

\renewcommand\tilde[1]{\mkern1mu\widetilde{\mkern-1mu#1}}

\newcommand\ii{\mathrm{i}}
\newcommand\ee{\mathrm{e}}
\newcommand\dd{\mathrm{d}}
\newcommand\ppi{\mathrm{\pi}}
\newcommand\tr{\mathsf{{\scriptscriptstyle T}}}
\newcommand\Tr{\operatorname{Tr}}
\newcommand\op[1]{\operatorname{#1}}
\newcommand\diag{\operatorname{diag}}
\newcommand\Lie{\mathrm{\mathscr{L}}}

\newcommand\mfrac[2]{\frac{\raisebox{-0.45ex}{\scalebox{0.9}{#1}}}{\raisebox{0.4ex}{\scalebox{0.9}{#2}}}}
\newcommand\mbinom[2]{\Big(\begin{array}{c}
 #1\\[-0.75ex]
 #2 
\end{array}\Big)}

\newcommand\QED{\mbox{{\color{red}\qedhere}}}

\newcommand\tudu[4]{?{\mbox{\ensuremath{#1}}}^{#2}{}_{#3}{}^{#4}?}
\newcommand\tdud[4]{?{\mbox{\ensuremath{#1}}}{}_{#2}{}^{#3}{}_{#4}?}
\newcommand\tud[3]{?{\mbox{\ensuremath{#1}}}^{#2}{}_{#3}?}
\newcommand\tdu[3]{?{\mbox{\ensuremath{#1}}}{}_{#2}{}^{#3}?}

\newcommand\repEq[2]{\repEqq{#1}{#2}}

\newcommand\Aeq[1]{\tag*{(#1)}}
\newcommand\PSeq[1]{\tag*{\llap{P\,}(#1)}}
\newcommand\BJeq[1]{\tag*{\llap{B\,}(#1)}}

\newcommand\ixA{a}
\newcommand\ixB{b}
\newcommand\ccVar{\mathcal{C}}

\newcommand\lidx[1]{\ ^{(#1)}\!}

\ifColors
  \newcommand\gSector[1]{{\color{blue}#1}}
  \newcommand\fSector[1]{{\color{red}#1}}
  \newcommand\hSector[1]{{\color{green}#1}}
  \newcommand\sSector[1]{{\color{green}#1}}
  \newcommand\mSector[1]{{\color{magenta}#1}}
  \newcommand\hrColor[1]{#1}
  \newcommand\VColor[1]{{\color{magenta}#1}}
  \newcommand\KColor[1]{{\color{teal}#1}}
  \newcommand\KVColor[1]{{\color{purple}#1}}
  \newcommand\qvf{{\color{orange}\xi}}
  \newcommand\betaSum{{\color{orange}m^{d}{\textstyle \sum_{n}}\beta_{n}}}
  \newcommand\betaSumL{{\color{orange}m^{d}{\textstyle {\displaystyle \sum_{n}}}\beta_{n}}}
\else
  \newcommand\gSector[1]{{#1}}
  \newcommand\fSector[1]{{#1}}
  \newcommand\hSector[1]{{#1}}
  \newcommand\sSector[1]{{#1}}
  \newcommand\mSector[1]{{#1}}
  \newcommand\hrColor[1]{{#1}}
  \newcommand\VColor[1]{{#1}}
  \newcommand\KColor[1]{{#1}}
  \newcommand\KVColor[1]{{#1}}
  \newcommand\qvf{\xi}
  \newcommand\betaSum{m^{d}{\textstyle \sum_{n}}\beta_{n}}
  \newcommand\betaSumL{m^{d}{\displaystyle \sum_{n}}\beta_{n}}
\fi

\newcommand\gMet{\gSector g}
\newcommand\gSp{\gSector{\gamma}}
\newcommand\gLapse{\gSector N}
\newcommand\gShift{\gSector N}
\newcommand\gShiftVec{\gSector{\vec{N}}}
\newcommand\gK{\gSector K}
\newcommand\gE{\gSector e}
\newcommand\gD{\gSector D}
\newcommand\gR{\gSector R}
\newcommand\gCS{\gSector{\Gamma}}
\newcommand\gVse{\gSector{V_{g}}}
\newcommand\gTse{\gSector{T_{g}}}
\newcommand\gEinst{\gSector{G_{g}}}
\newcommand\gRicci{\gSector{R_{g}}}
\newcommand\gCC{\gSector{\mathcal{C}}}
\newcommand\gCE{\gSector{\mathcal{E}}}
\newcommand\gCD{\gSector{\nabla}}

\newcommand\fMet{\fSector f}
\newcommand\fSp{\fSector{\varphi}}
\newcommand\fLapse{\fSector M}
\newcommand\fShift{\fSector M}
\newcommand\fShiftVec{\fSector{\vec{M}}}
\newcommand\fK{\fSector{\tilde{K}}}
\newcommand\fE{\fSector m}
\newcommand\fD{\fSector{\tilde{D}}}
\newcommand\fR{\fSector{\tilde{R}}}
\newcommand\fCS{\fSector{\tilde{\Gamma}}}
\newcommand\fVse{\fSector{V_{f}}}
\newcommand\fTse{\fSector{T_{f}}}
\newcommand\fEinst{\fSector{G_{f}}}
\newcommand\fRicci{\fSector{R_{f}}}
\newcommand\fCC{\fSector{\widetilde{\mathcal{C}}}}
\newcommand\fCE{\fSector{\widetilde{\mathcal{E}}}}
\newcommand\fCD{\fSector{\widetilde{\nabla}}}

\newcommand\gKappa{\gSector{\kappa_{g}}}
\newcommand\gKappainv{\gSector{\kappa_{g}^{-1}}}
\newcommand\Mg{\gSector{M_{g}^{d-2}}}

\newcommand\fKappa{\fSector{\kappa_{f}}}
\newcommand\fKappainv{\fSector{\kappa_{f}^{-1}}}
\newcommand\Mf{\fSector{M_{f}^{d-2}}}

\newcommand\grho{\gSector{\rho}}
\newcommand\gjota{\gSector j}
\newcommand\gJota{\gSector J}

\newcommand\frho{\fSector{\tilde{\rho}}}
\newcommand\fjota{\fSector{\tilde{j}}}
\newcommand\fJota{\fSector{\tilde{J}}}

\newcommand\gAlpha{\gSector{\alpha}}
\newcommand\gBeta{\gSector{\beta}}
\newcommand\gEA{\gSector A}
\newcommand\gEB{\gSector B}
\newcommand\fAlpha{\fSector{\tilde{\alpha}}}
\newcommand\fBeta{\fSector{\tilde{\beta}}}
\newcommand\fEA{\fSector{\tilde{A}}}
\newcommand\fEB{\fSector{\tilde{B}}}

\newcommand\sEtau{\mSector{\tau}}
\newcommand\sESigma{\mSector{\Sigma}}
\newcommand\sER{\mSector R}

\newcommand\Proj{\operatorname{\perp}}
\newcommand\gProj{\gSector{\operatorname{\perp}_{g}}}
\newcommand\fProj{\fSector{\operatorname{\perp}_{f}}}
\newcommand\hProj{\hSector{\operatorname{\perp}}}
\newcommand\prho{\boldsymbol{\rho}}
\newcommand\pjota{\boldsymbol{j}}
\newcommand\pJota{\boldsymbol{J}}

\newcommand\sgn{\gSector{\mathsfit{n}{\mkern1mu}}}
\newcommand\sgD{\gSector{\mathcal{D}}}
\newcommand\sgQ{\gSector{\mathcal{Q}}}
\newcommand\sgV{\gSector{\mathcal{V}}}
\newcommand\sgU{\gSector{\mathcal{U}}}
\newcommand\sgB{\gSector{\mathcal{B}}}

\newcommand\sfn{\fSector{\tilde{\mathsfit{n}}{\mkern1mu}}}
\newcommand\sfD{\fSector{\widetilde{\mathcal{D}}}}
\newcommand\sfQ{\fSector{\widetilde{\mathcal{Q}}}}
\newcommand\sfV{\fSector{\widetilde{\mathcal{V}}}}
\newcommand\sfU{\fSector{\widetilde{\mathcal{U}}}}
\newcommand\sfB{\fSector{\widetilde{\mathcal{B}}}}

\newcommand\sgW{\gSector{\mathcal{W}}}
\newcommand\sgQU{\gSector{(\mathcal{Q\fSector{{\scriptstyle \widetilde{U}}}})}}

\newcommand\sfW{\fSector{\tilde{\mathcal{W}}}}
\newcommand\sfQU{\fSector{(\mathcal{\widetilde{Q}\gSector{{\scriptstyle U}}})}}

\newcommand\hMet{\hSector h}
\newcommand\hSp{\hSector{\chi}}
\newcommand\hLapse{\hSector H}
\newcommand\hShift{\hSector q}
\newcommand\hShiftVec{\hSector q}
\newcommand\hCC{\hSector{\bar{\mathcal{C}}}}

\newcommand\sLs{\sSector{\hat{\Lambda}}}
\newcommand\sLt{\sSector{\lambda}}
\newcommand\sLtinv{\sSector{\lambda^{-1}}}
\newcommand\sLv{\hSector v}
\newcommand\sLp{\hSector p}
\newcommand\sRs{\hSector{\hat{R}}}
\newcommand\sRbar{\hSector{\bar{R}}}

\newcommand\sI{\sSector{\hat{I}}}
\newcommand\sEta{\sSector{\hat{\delta}}}

\ifColors

  \newcommand\signV{\,\VColor{\stackrel{{\scriptscriptstyle (V)}}{-}}\,}
  \newcommand\usignV{\VColor{\stackrel{{\scriptscriptstyle (V)}}{-}}\,}
  \newcommand\isignV{\,\VColor{\stackrel{{\scriptscriptstyle (V)}}{+}}\,}
  \newcommand\uisignV{\VColor{\stackrel{{\scriptscriptstyle (V)}}{+}}\,}

  \newcommand\signK{\,\KColor{\stackrel{{\scriptscriptstyle (K)}}{-}}\,}
  \newcommand\usignK{\KColor{\stackrel{{\scriptscriptstyle (K)}}{-}}\,}
  \newcommand\isignK{\,\KColor{\stackrel{{\scriptscriptstyle (K)}}{+}}\,}
  \newcommand\uisignK{\KColor{\stackrel{{\scriptscriptstyle (K)}}{+}}\,}

  \newcommand\signKV{\,\KVColor{\stackrel{{\scriptscriptstyle (KV)}}{+}}\,}
  \newcommand\usignKV{\KVColor{\stackrel{{\scriptscriptstyle (KV)}}{+}}\,}
  \newcommand\isignKV{\,\KVColor{\stackrel{{\scriptscriptstyle (KV)}}{-}}\,}
  \newcommand\uisignKV{\KVColor{\stackrel{{\scriptscriptstyle (KV)}}{-}}\,}

\else

  \newcommand\signV{\,-\,}
  \newcommand\usignV{-\,}
  \newcommand\isignV{\,+\,}
  \newcommand\uisignV{}
  
  \newcommand\signK{\,-\,}
  \newcommand\usignK{-\,}
  \newcommand\isignK{\,+\,}
  \newcommand\uisignK{}
  
  \newcommand\signKV{\,+\,}
  \newcommand\usignKV{}
  \newcommand\isignKV{\,-\,}
  \newcommand\uisignKV{-\,}

\fi

\newcommand\hrD{\hrColor D}
\newcommand\hrQ{\hrColor Q}
\newcommand\hrn{\hrColor n}
\newcommand\hrDn{\hrColor{Dn}}
\newcommand\hrx{\hrColor x}
\newcommand\hrV{\hrColor V}
\newcommand\hrU{\hrColor U}
\newcommand\hrVbar{\hrColor{\bar{V}}}
\newcommand\hrWbar{\hrColor{\bar{W}}}
\newcommand\hrSV{\hrColor S}
\newcommand\hrUtilde{\hrColor{\tilde{U}}}
\newcommand\hrVubar{\hrColor{\underbar{V}}}

\fi  


\newcommand{\myReleaseInfo}{~\\~}

\newcommand{\myTitle}{Causal propagation of constraints in bimetric
relativity in standard 3+1 form}

\newcommand{\myAbstract}{The goal of this work was to investigate
the propagation of the constraints in the ghost-free bimetric theory
where the evolution equations are in standard 3+1 form. It is established
that the constraints evolve according to a first-order symmetric hyperbolic
system whose characteristic cone consists of the null cones of the
two metrics. Consequently, the constraint evolution equations are
well-posed, and the constraints stably propagate.}

\newcommand{\myKeywords}{\bgroup\small Modified gravity, Ghost-free
bimetric theory, Constraints, Initial-value problem\egroup}

\title{\myTitle}

\author{Mikica Kocic}

\affiliation{
  Department of Physics \& The Oskar Klein Centre,\\
  Stockholm University, AlbaNova University Centre,
  SE-106 91 Stockholm
}

\email{mikica.kocic@fysik.su.se}

\hypersetup{
  pdftitle=\myTitle,
  pdfauthor=Mikica Kocic,
  pdfsubject=Hassan-Rosen ghost-free bimetric theory,
  pdfkeywords={Modified gravity, Ghost-free bimetric theory,
    Constraints, Initial-value problem}
}

\emitFrontMatter


\section{Introduction}

We study the propagation of the constraints in the Hassan-Rosen (HR)
ghost-free bimetric theory \cite{Hassan:2011zd,Hassan:2011ea,Hassan:2017ugh,Hassan:2018mbl}
where the evolution equations are expressed in standard 3+1 form \cite{Kocic:2018ddp}.
The HR theory is a nonlinear theory of two interacting classical spin-2
fields, which is closely related to de Rham\textendash Gabadadze\textendash Tolley
(dRGT) massive gravity \cite{deRham:2010ik,deRham:2010kj,Hassan:2011hr}.
 Like in general relativity (GR), the unphysical degrees of freedom
in the HR theory are necessarily eliminated by constraints. Also as
in GR, one of the difficulties encountered when treating the initial
value problem (IVP) is the fact that the bimetric theory is a constrained
system.

Suppose that we have reduced the bimetric field equations and posed
the IVP for the HR theory. Besides the evolution equations, the IVP
setup will comprise a set of constraints $\{\ccVar_{n}\}$ that obey
evolution equations of the form $\partial_{t}\ccVar_{n}=F_{n}(\ccVar_{1},\ccVar_{2},\dots)$
where $\partial_{t}\ccVar_{n}=0$ if all $\ccVar_{n}$ vanish. Then,
for analytic initial data,  if the constraints exactly vanish on
the initial manifold, they will be vanishing at all times due to the
Cauchy\textendash Kovalevskaya theorem.  

Nevertheless, analytic functions are fully determined by the values
on a small open set, which badly fits with causality requirements
and the notion of the domain of dependence. Also, on physical grounds,
 assuming analytic initial data is too restrictive because of the
diffeomorphism invariance of the theory and the smooth structure of
the manifold. In fact, the real problem is in the requirement `exactly
vanish' which is unobtainable since the physical quantities are a
priori given with some uncertainty. 

Because the physical initial data cannot be freely specified, even
infinitesimal variations that violate the constraints can lead to
significantly different values at subsequent times. As a result, the
continuous dependence on initial data will be corrupted, destroying
the well-posedness \cite{Hadamard:1902a} of a mathematical problem
that is to correspond to physical reality. Therefore, it is important
to show that the constraints propagate in a stable manner for the
given IVP setup in bimetric theory. This is in particular relevant
to the problem of evolution in numerical relativity, if the constraints
are solved only to get the initial data. Otherwise, if the constraint
evolution equations are not well-posed, the unphysical modes will
not be bounded but uncontrollably amplified during the free evolution.
\newpage

The causal propagation of the constraints in general relativity where
the evolution equations are in standard 3+1 form as formulated by
York \cite{York:1979aa} was established by Frittelli in \cite{Frittelli:1996nj}.
A similar procedure is followed in this work. The starting point of
the performed analysis comprises the bimetric equations in standard
3+1 form obtained in \cite{Kocic:2018ddp}. 

This paper is organized as follows. In the rest of the introduction,
we review the HR field equations and the bimetric space-plus-time
split \`{a} la York. In section 2, after revisiting how the propagation
of the constraints works in GR, we show that the bimetric constraints
evolve according to a first-order symmetric hyperbolic system whose
characteristic cone consists of the null cones of the two metrics.
The paper ends with a summary and outlook.

\subsection{Bimetric field equations}

\label{sec:bim-fe}

Let $\gMet$ and $\fMet$ be two metric fields on a $d$-dimensional
manifold coupled through a ghost-free bimetric potential \cite{deRham:2010ik,deRham:2010kj,Hassan:2011vm},
\begin{equation}
V(S)\coloneqq\usignV\betaSum\,e_{n}(S),\qquad S\coloneqq(\gMet^{-1}\fMet)^{1/2}.\label{eq:V}
\end{equation}
Here, $(\gMet^{-1}\fMet)^{1/2}$ denotes the principal square root
of the operator $\gMet^{\mu\rho}\fMet_{\rho\nu}$, and $e_{n}(S)$
are the elementary symmetric polynomials \cite{macdonald:1998a},
the scalar invariants of $S$ which can be expressed,
\begin{equation}
e_{n\ge1}(S)\coloneqq\tud S{[\mu_{1}}{\mu_{1}}\tud S{\mu_{2}}{\mu_{2}}\cdots\tud S{\mu_{n}]}{\mu_{n}},\qquad e_{0}(S)\coloneqq1.
\end{equation}
In particular, $e_{1}(S)=\Tr S$, $e_{d}(S)=\det S$, and $e_{n>d}(S)=0$.
The potential  is parametrized by the set of dimensionless real constants
$\{\beta_{n}\}$ and an overall mass scale $m$. The algebraic form
of the potential is due to the necessary condition for the absence
of ghosts \cite{Creminelli:2005qk} where the dynamics of each metric
is given by a separate Einstein-Hilbert term in the action \cite{Hassan:2011zd}.
The principal branch of $S$ ensures an unambiguous definition of
the theory \cite{Hassan:2017ugh}.

Since we are interested in the partial differential equations (PDE)
governing the theory, we start from the locally given bimetric field
equations,\bSe\label{eq:bim-eom}
\begin{alignat}{2}
\gEinst & =\gKappa\gVse+\gKappa\gTse, & \qquad\gVse & \coloneqq\usignV\betaSum\,Y_{n}(S),\\[1ex]
\fEinst & =\fKappa\fVse+\fKappa\fTse, & \qquad\fVse & \coloneqq\usignV\betaSum\,Y_{d-n}(S^{-1}),
\end{alignat}
\eSe where $\gEinst$ and $\fEinst$ are the Einstein tensors of
the two metrics, $\gKappa$ and $\fKappa$ are two different gravitational
constants, $\gTse$ and $\fTse$ are the stress-energy tensors of
the matter fields each minimally coupled to a different sector, and
$\gVse$ and $\fVse$ are the effective stress-energy contributions
of the bimetric potential (\ref{eq:V}). The function $Y_{n}(S)$
in (\ref{eq:bim-eom}) encapsulates the variation of the bimetric
potential with respect to the metrics (note that  $Y_{n\ge d}=0$),
\begin{gather}
Y_{n}(S)\coloneqq\sum_{k=0}^{n}(-1)^{n+k}e_{k}(S)\,S^{n-k}=\frac{\partial e_{n+1}(S)}{\partial S^{\tr}}.\label{eq:Yn-def}
\end{gather}
Importantly, the effective stress-energy tensors satisfy the following
identities \cite{Hassan:2014vja,Damour:2002ws},\bSe\label{eq:bim-ids}
\begin{align}
\sqrt{-\gMet}\,\tud{\gVse}{\mu}{\nu}+\sqrt{-\fMet}\,\tud{\fVse}{\mu}{\nu}-\sqrt{-\gMet}\,V\,\tud{\delta}{\mu}{\nu} & =0,\label{eq:id-alg2}\\[1ex]
\sqrt{-\gMet}\,\gCD_{\mu}\gVse{}^{\mu}{}_{\nu}+\sqrt{-\fMet}\,\fCD_{\mu}\fVse{}^{\mu}{}_{\nu} & =0,\label{eq:id-damour}
\end{align}
\eSe where $\gCD_{\mu}$ and $\fCD_{\mu}$ are the covariant derivatives
compatible with $\gMet$ and $\fMet$, respectively. Assuming that
the matter conservation laws hold $\gCD_{\mu}\tud{\gTse}{\mu}{\nu}=0$
and $\fCD_{\mu}\tud{\fTse}{\mu}{\nu}=0$, the field equations (\ref{eq:bim-eom})
imply the \emph{bimetric conservation law},
\begin{equation}
\gCD_{\mu}\tud{\gVse}{\mu}{\nu}=0,\qquad\fCD_{\mu}\tud{\fVse}{\mu}{\nu}=0.\label{eq:bianchi}
\end{equation}
The two equations in (\ref{eq:bianchi}) are not independent according
to the differential identity (\ref{eq:id-damour}).

\subsection{\nPlusOne{} splitting}

In GR, the kinematical and dynamical parts of a metric field can be
isolated using the \nPlusOne{} formalism \cite{Arnowitt:1962hi,York:1979aa}.
A similar procedure can be applied to the bimetric theory. However,
one must also take into account: (i) the simultaneous \nPlusOne{}
decomposition in both sectors, (ii) the parametrization of the metric
fields which does not corrupt the separation between the kinematical
and the dynamical parts, and (iii) the bimetric conservation law. 

For the first, the existence of a common spacelike hypersurface with
respect to both metrics is related to the existence of the real square
root $S$ by the theorem from \cite{Hassan:2017ugh}. For the second,
the parametrization can be based on the geometric mean metric $\hMet=\gMet S=\fMet S^{-1}$.
As shown in \cite{Kocic:2018ddp}, such a parametrization covers all
possible metric configurations which can have the real principal square
root $S$. Then, the projection of the bimetric conservation law is
straightforward \cite{Kocic:2018ddp}, giving the correct number of
truly dynamical degrees of freedom.

Let us consider a particular metric sector, say $\gMet$. A foliation
of the spacetime into a family of spacelike hypersurfaces $\{\Sigma\}$
is assumed with the timelike unit normal $\vec{n}$ on the slices
such that $n_{\mu}n^{\mu}=-1$ with respect to $\gMet$. The geometry
is being projected onto the spacelike slices using the operator,
\begin{equation}
\tud{\Proj}{\mu}{\nu}\coloneqq\tud{\delta}{\mu}{\nu}+n^{\mu}n_{\nu}.
\end{equation}
Any symmetric tensor field $X_{\mu\nu}$ can be decomposed into the
perpendicular projection $\prho$, the mixed projection $\pjota$,
and the full projection $\pJota$ of $X$ onto $\Sigma$,
\begin{equation}
\prho[X]\coloneqq n^{\mu}X_{\mu\nu}n^{\nu},\quad\pjota[X]_{\mu}\coloneqq-\tud{\Proj}{\nu}{\mu}X_{\nu\sigma}n^{\sigma},\quad\pJota[X]_{\mu\nu}\coloneqq\tud{\Proj}{\rho}{\mu}X_{\rho\sigma}\tud{\Proj}{\sigma}{\nu},\label{eq:proj-ops}
\end{equation}
such that,
\begin{equation}
X_{\mu\nu}=\prho[X]\,n_{\mu}n_{\nu}\,+\,n_{\mu}\pjota[X]_{\nu}+\pjota[X]_{\mu}n_{\nu}+\pJota[X]_{\mu\nu},\label{eq:proj-X}
\end{equation}
with the trace,
\begin{equation}
\tud X{\mu}{\mu}=\gMet^{\mu\nu}X_{\mu\nu}=\gSp^{ij}\pJota[X]_{ij}-\prho[X]=\tud{\pJota[X]}ii-\prho[X].\label{eq:proj-X-trace}
\end{equation}
For $X=\gMet$, we have $\prho=-1$, $\pjota=0$, and the metric induced
on the spatial slices reads,
\begin{equation}
\gSp_{\mu\nu}\coloneqq\pJota[\gMet]_{\mu\nu}=\gMet_{\mu\nu}+n_{\mu}n_{\nu}.\label{eq:kd-7}
\end{equation}
This implies $\gMet_{\mu\nu}=-n_{\mu}n_{\nu}+\gSp_{\mu\nu}$. Now,
in a suitable chart $x^{\mu}=(t,x^{i})$ where,
\begin{equation}
n_{\mu}=\big(-\gLapse,0\big),\qquad n^{\mu}=\big(\gLapse^{-1},-\gLapse^{-1}\gShift{}^{i}\big),
\end{equation}
the metric can be written,
\begin{equation}
\gMet=-\gLapse^{2}\dd t^{2}+\gSp_{ij}\big(\dd x^{i}+\gShift{}^{i}\dd t\big)\big(\dd x^{j}+\gShift{}^{j}\dd t\big).\label{eq:kd-6}
\end{equation}
Here, $\gLapse$ and $\gShift^{i}$ are the standard lapse function
and shift vector, respectively. The shift $\gShift{}^{i}$ is a purely
spatial vector field, and $\gSp_{ij}$ is a spatial Riemannian metric
whose inverse is obtained through $\gSp^{ik}\gSp_{kj}=\delta_{j}^{i}$. 

The bimetric field equations (\ref{eq:bim-eom}) can be decomposed
employing the projections (\ref{eq:proj-ops}) with the help of the
Gauss\textendash Codazzi\textendash Mainardi equations. Let $\gK_{ij}$
be the extrinsic curvature of the slices with respect to $\gMet$,
\begin{equation}
\gK_{ij}\coloneqq\usignK\mfrac 12\,\Lie_{\vec{n}}\gSp_{ij},\qquad\gK\coloneqq\gSp^{ij}\gK_{ij},\label{eq:K-def}
\end{equation}
where $\Lie_{\vec{n}}$ denotes the Lie derivative along the vector
field $\vec{n}$. The perpendicular and the mixed projection of the
Einstein tensor $\gEinst$ reads,\bSe\label{eq:rhoj-G}
\begin{alignat}{2}
\prho[\gEinst]_{\hphantom{i}} & =\,\, & \gLapse^{2}\gEinst^{00} & =\mfrac 12\big(\gR+\gK^{2}-\gK_{ij}\gK^{ij}\big),\\
\pjota[\gEinst]_{i} & =\,\, & -\gLapse\tud{\gEinst}0i & =\ifShowSigns\isignK\big(\gD_{j}\tud{\gK}ji-\gD_{i}\gK\big)\else\uisignK\gD_{j}\tud{\gK}ji\signK\gD_{i}\gK\fi,
\end{alignat}
\eSe where $\gR=\gSp^{ij}\gR_{ij}$ is the trace of the Ricci tensor
$\gR_{ij}$ which is defined using the spatial covariant derivatives
$\gD_{i}$ compatible with $\gSp_{ij}$. The expressions (\ref{eq:rhoj-G})
will constitute the constraint equations in the $\gMet$-sector. Following
York \cite{York:1979aa}, the evolution equations in standard \nPlusOne{}
form are obtained by projecting the equations of motion,
\begin{equation}
\tud{\gRicci}{\mu}{\nu}=\gKappa\tud{(\gVse+\gTse)}{\mu}{\nu}-\frac{1}{d-2}\gKappa\tud{(\gVse+\gTse)}{\sigma}{\sigma}\tud{\delta}{\mu}{\nu},
\end{equation}
which are based on the $d$-dimensional Ricci tensor. The full spatial
projection of $\gRicci$ gives,
\begin{align}
\pJota[\gRicci]_{ij} & =\uisignK\gLapse^{-1}\big(\partial_{t}\gK_{ij}-\Lie_{\gShiftVec}\gK_{ij}\big)-\gLapse^{-1}\gD_{i}\gD_{j}\gLapse+\gR_{ij}-2\gK_{ik}\tud{\gK}kj+\gK\gK_{ij}.\label{eq:proj-R}
\end{align}
Combining (\ref{eq:rhoj-G}) and (\ref{eq:proj-R}) with the stress-energy
tensor projections,
\begin{equation}
\grho\coloneqq\prho[\gVse+\gTse],\qquad\gjota\coloneqq\pjota[\gVse+\gTse],\qquad\gJota\coloneqq\pJota[\gVse+\gTse],
\end{equation}
we get the constraint and the evolution equations in the $\gMet$-sector,
respectively.

The similar expressions are established in the $\fMet$-sector where
the geometry is projected with respect to the timelike unit normal
$n_{\fMet}^{\mu}=\big(\fLapse^{-1},-\fLapse^{-1}\fShift{}^{i}\big)$
relative to $\fMet$ such that,
\begin{equation}
\fMet=-\fLapse^{2}\dd t^{2}+\fSp_{ij}\big(\dd x^{i}+\fShift{}^{i}\dd t\big)\big(\dd x^{j}+\fShift{}^{j}\dd t\big).\label{eq:kd-6-1}
\end{equation}
The rest of the variables in the $\fMet$-sector are denoted by tildes:
the extrinsic curvature $\fK_{ij}$, the spatial Ricci tensor $\fR_{ij}$,
and the spatial covariant derivative $\fD_{i}$. The stress-energy
tensor projections in the $\fMet$-sector are (obtained using the
timelike unit normal $\vec{n}_{\fMet}$),
\begin{equation}
\frho\coloneqq\prho[\fVse+\fTse],\qquad\fjota\coloneqq\pjota[\fVse+\fTse],\qquad\fJota\coloneqq\pJota[\fVse+\fTse].
\end{equation}
Note that one can always find a common spacelike hypersurface with
respect to both metrics, if a real principal square root $S$ exists
\cite{Hassan:2017ugh}.  Moreover, the parametrization \cite{Kocic:2018ddp}
will give all possible metric configurations that have a real principal
square root $S$. At the end, the space-plus-time split will comprise
the following set of \nPlusOne{} variables,
\begin{equation}
\{\sgn,\sgD,\sgB,\sgV,\sgU,\sgQ,\sfn,\sfD,\sfB,\sfV,\sfU,\sfQ,\sLt\}.\label{eq:Np1-vars}
\end{equation}
These variables do not depend on the shift of the mean metric $\hMet$
and the lapses $\gLapse$ and $\fLapse$; they are quoted in appendix
\ref{app:bim-vars} as their form is not important for further analysis.

The evolution equations for the dynamical pairs $(\gSp_{ij},\gK_{ij})$
and $(\fSp_{ij},\fK_{ij})$ reads,\bSe\label{eq:g-evol}
\begin{align}
\partial_{t}\gSp_{ij} & =\Lie_{\gShiftVec}\gSp_{ij}\signK2\gLapse\gK_{ij},\\
\partial_{t}\gK_{ij} & =\Lie_{\gShiftVec}\gK_{ij}\signK\gD_{i}\gD_{j}\gLapse\isignK\gLapse\,\big[\gR_{ij}-2\gK_{ik}\gK^{k}{}_{j}+\gK\gK_{ij}\big]\nonumber \\
 & \qquad\quad\signK\gLapse\gKappa\Big\{\,\gSp_{ik}\tud{\gJota}kj-\frac{1}{d-2}\gSp_{ij}(\gJota-\grho)\,\Big\},\\
\partial_{t}\fSp_{ij} & =\Lie_{\fShiftVec}\fSp_{ij}\signK2\fLapse\fK_{ij},\\
\partial_{t}\fK_{ij} & =\Lie_{\fShiftVec}\fK_{ij}\signK\fD_{i}\fD_{j}\fLapse\isignK\fLapse\,\big[\fR_{ij}-2\fK_{ik}\fK^{k}{}_{j}+\fK\fK_{ij}\big]\\
 & \qquad\quad\signK\fLapse\fKappa\Big\{\,\fSp_{ik}\tud{\fJota}kj-\frac{1}{d-2}\fSp_{ij}(\fJota-\frho)\,\Big\}.
\end{align}
\eSe The scalar and vector constraint equations are,\bSe\label{eq:g-cc}\bgroup
\begin{alignat}{3}
\gCC & \coloneqq\mfrac 12(\gR+\gK^{2}-\gK_{ij}\gK^{ij}) & \, & \,- & \,\gKappa\,\grho & =0,\label{eq:g-cc-scalar}\\
\fCC & \coloneqq\mfrac 12(\fR+\fK^{2}-\fK_{ij}\fK^{ij}) & \, & \,- & \,\fKappa\,\frho & \,=0,\label{eq:f-cc-scalar}\\
\gCC_{i} & \coloneqq\gD_{k}\gK^{k}{}_{i}-\gD_{i}\gK & \, & \signK & \gKappa\,\gjota_{i} & =0,\label{eq:g-cc-vector}\\
\fCC_{i} & \coloneqq\fD_{k}\fK^{k}{}_{i}-\fD_{i}\fK & \, & \signK & \fKappa\,\fjota_{i} & =0.\label{eq:f-cc-vector}
\end{alignat}
\egroup\eSe The full spatial projection of $\gVse+\gTse$ and $\fVse+\fTse$
are given by,\bSe
\begin{align}
\tud{\gJota}kj & =\big\llbracket\,\sgV\sI-\sgQU+\fLapse\gLapse^{-1}\sgU\,\tud{\big\rrbracket}kj+\pJota[\gTse\tud ]kj,\\
\tud{\fJota}kj & =\frac{\sqrt{\gSp}}{\sqrt{\fSp}}\big\llbracket\,\sfV\sI-\sgQU+\gLapse\fLapse^{-1}\sfU\,\tud{\big\rrbracket}kj+\pJota[\fTse\tud ]kj,
\end{align}
\eSe where the traces of $\gVse+\gTse$ and $\fVse+\fTse$ are respectively
$\gJota-\grho$ and $\fJota-\frho$ by using (\ref{eq:proj-X-trace}).
The perpendicular and mixed projections reads,\bSe\bgroup
\begin{alignat}{2}
\grho & =\uisignV\betaSumL\,e_{n}(\sgB)+\prho[\gTse]\,, & \gjota_{i} & =\pjota[\gVse]_{i}+\pjota[\gTse]_{i}\,,\\
\frho & =\uisignV\frac{\sqrt{\gSp}}{\sqrt{\fSp}}\betaSumL\sLt e_{d-1}(\sfD)+\prho[\fTse]\,, & \fjota_{i} & =\pjota[\fVse]_{i}+\pjota[\fTse]_{i}\,,\\
\pjota[\gVse]_{i} & =-\gSp_{ij}\tud{\sgQU}jk\sgn^{k}\,, & \qquad0 & =\sqrt{\gSp}\,\pjota[\gVse]_{i}+\sqrt{\fSp}\,\pjota[\fVse]_{i}\,.\label{eq:f-cc-j-def}
\end{alignat}
\egroup\eSe 

In the \nPlusOne{} decomposition, we also have to assume specifically
that (i) the matter conservation laws $\gCD_{\mu}\tud{\gTse}{\mu}{\nu}=0$
and $\fCD_{\mu}\tud{\fTse}{\mu}{\nu}=0$ hold, and (ii) the conservation
law for the bimetric potential $\gCD_{\mu}\tud{\gVse}{\mu}{\nu}=0$
holds. The projection of $\gCD_{\mu}\tud{\gVse}{\mu}{\nu}=0$ gives
the \nPlusOne{} form of the conservation law for the ghost-free bimetric
potential \cite{Kocic:2018ddp},
\begin{equation}
\tud{\sgU}ij\Big(\gD_{i}\sgn^{j}\signK\tud{\gK}ji\Big)+\tud{\sfU}ij\Big(\fD_{i}\sfn^{j}\isignK\tud{\fK}ji\Big)-\gD_{i}\big[\tud{\sgU}ij\sgn^{j}\big]=0.\label{eq:bim-prop-constr}
\end{equation}
The equation (\ref{eq:bim-prop-constr}) is the same as the so-called
\emph{secondary constraint} obtained using the Hamiltonian formalism
\cite{Hassan:2018mbl}. The existence of (\ref{eq:bim-prop-constr})
is essential for removing the unphysical (ghost) modes. The counting
of the degrees of freedom for the system (\ref{eq:g-evol})\textendash (\ref{eq:bim-prop-constr})
is given in Table 2 in \cite{Kocic:2018ddp}. Importantly, the lapses
of $\gMet$ and $\fMet$, and the shift of the geometric mean $\hMet$
are absent from the constraint equations and from the projected conservation
laws.

\section{Propagation of constraints}

We first consider the propagation of the constraints in general relativity.
The GR equations are in fact the same as the HR equations if the bimetric
interaction is turned off, $V(S)=0$. Then $\gVse=\fVse=0$, and the
two metric sectors decouple.

Let us assume that we have prepared GR data that satisfies the constraint
equations on the initial spatial hypersurface at $t=0$. We then evolve
the system by solving the evolution equations for $(\gSp_{ij},\gK_{ij})$,
which have to satisfy the constraint equations on each spatial hypersurface
of $t>0$. The constraint equations are satisfied due the contracted
Bianchi identity provided that the conservation laws $\gCD_{\mu}\tud{\gTse}{\mu}{\nu}=0$
hold, which is automatically fulfilled if one solves the field equations. 

In the \nPlusOne{} decomposition, however, we need to assume \emph{specifically}
that $\gCD_{\mu}\tud{\gTse}{\mu}{\nu}=0$ holds because we do not
demand that the complete set of the field equations is satisfied,
but only their dynamical part \cite{Gourgoulhon:2012trip,Alcubierre:2012intro,Shibata:2015nr,Bona:2009el},
\begin{equation}
\tud{\gCE}ij=0,\qquad\tud{\gCE}ij\coloneqq\pJota[\gRicci\tud ]ij-\frac{1}{d-2}\Big(J-\rho\Big)\tud{\delta}ij,\label{eq:E-efe}
\end{equation}
where $\rho=\prho[\gTse]$, $j_{i}=\pjota[\gTse]_{i}$, $\tud Jij=\pJota[\gTse\tud ]ij$,
and $J=\tud Jkk$. In this context, the matter energy $\rho$ and
momentum density $j$ must obey the equation $\gCD_{\mu}\tud{\gTse}{\mu}{\nu}=0$.
A straightforward manipulation \cite{Gourgoulhon:2012trip,Shibata:1995we}
yields the following projections of $\gCD_{\mu}\tud{\gTse}{\mu}{\nu}=0$,\bSe\label{eq:T-evol}
\begin{align}
\partial_{t}\rho & =\Lie_{\gShiftVec}\rho-\gLapse\gD_{i}j^{i}-2j^{i}\gD_{i}\gLapse\isignK\gLapse\gK\rho\isignK\gLapse\tud{\gK}ji\tud Jij,\\
\partial_{t}j_{i} & =\Lie_{\gShiftVec}j_{i}-\gD_{j}\big[\gLapse\tud Jji\big]-\rho\gD_{i}\gLapse\isignK\gLapse\gK j_{i}.
\end{align}
\eSe These are effectively the evolution equations of the matter
fields in the \nPlusOne{} decomposition. An important remark is that,
if the bimetric interaction is taken into account, the projection
of the bimetric conservation law $\gCD_{\mu}\tud{\gVse}{\mu}{\nu}=0$
yields the constraint equation (\ref{eq:bim-prop-constr}).

Now, thanks to the contracted Bianchi identity $\gCD_{\mu}\tud{\gEinst}{\mu}{\nu}=0$
and the fulfilled conservation laws, the projection of the divergence
of the field equations (\ref{eq:bim-eom}) is obtained similarly to
(\ref{eq:T-evol}) by a formal replacement (see also chapter 11 in
\cite{Gourgoulhon:2012trip}),
\begin{equation}
\rho\to\gCC,\quad j_{i}\to\gCC_{i},\quad\text{and}\quad\tud Jij\to J[\gEinst-\gKappa\gTse\tud ]ij=\tud{\gCE}ij+(\gCC-\gCE)\tud{\delta}ij.
\end{equation}
Here we used $\gCC$ and $\gCC_{i}$ to denote the violations of the
constraints rather than the constraints. As a result, we get the evolution
equations for the constraint violations,\bSe\label{eq:CC-evol}
\begin{align}
\partial_{t}\gCC & =\Lie_{\gShiftVec}\gCC-\gLapse\gD_{i}\gCC^{i}-2\gCC^{i}\gD_{i}\gLapse\isignK\gLapse\gK\gCC\isignK\gLapse\Big(\tud{\gK}ji\tud{\gCE}ij-\gK\gCE\Big),\\
\partial_{t}\gCC_{i} & =\Lie_{\gShiftVec}\gCC_{i}-\gD_{j}\big[\gLapse(\tud{\gCE}ji-\gCE\tud{\delta}ji)\big]-2\gCC\gD_{i}\gLapse-\gLapse\gD_{i}\gCC\isignK\gLapse\gK\gCC_{i}.
\end{align}
\eSe Assuming that the evolution equations hold, $\tud{\gCE}ij=0$,
we have,\bSe\label{eq:CC-evol-2}
\begin{align}
\partial_{t}\gCC & =\Lie_{\gShiftVec}\gCC-\gLapse\gD_{i}\gCC^{i}-2\gCC^{i}\gD_{i}\gLapse\isignK\gLapse\gK\gCC,\\
\partial_{t}\gCC_{i} & =\Lie_{\gShiftVec}\gCC_{i}-2\gCC\gD_{i}\gLapse-\gLapse\gD_{i}\gCC\isignK\gLapse\gK\gCC_{i}.
\end{align}
\eSe If the constraints are satisfied on the initial spatial hypersurface
at $t=0$, then $\gCC=0$, $\gCC_{i}=0$, and we have $\partial_{t}\gCC=0$
and $\partial_{t}\gCC_{i}=0$. Hence, provided that (\ref{eq:T-evol})
holds, the constraints are preserved for all $t>0$ by the dynamical
evolution, at least in the case when the initial data is analytic.
Nonetheless, the same conclusion can be deduced even for the smooth
initial data since the system (\ref{eq:CC-evol-2}) is in a symmetric
hyperbolic form \cite{Frittelli:1996nj}.

Note that the \nPlusOne{} evolution equations (\ref{eq:g-evol})
are not dynamically equivalent to the original Arnowitt\textendash Deser\textendash Misner
(ADM) equations \cite{Arnowitt:1962hi}. The source of the different
behavior is that the ADM equations come from the field equations written
in terms of the Einstein tensor $\gEinst$, while the evolution equations
(\ref{eq:g-evol}) are due to  York \cite{York:1979aa} derived from
the field equations expressed in terms of the Ricci tensor $\gRicci$.
 To see the difference, let us relate the ADM canonical momenta $\gSector{\pi}^{ij}$
conjugate to $\gSp_{ij}$ and the extrinsic curvature $\gK_{ij}$
by,
\begin{equation}
\gSector{\pi}^{ij}=\usignK\gKappa\sqrt{\gSp}\,\big(\gK^{ij}-\gK\gamma^{ij}\big).
\end{equation}
The ADM version \cite{Arnowitt:1959ah} of the evolution equation
for $\gK_{ij}$ becomes,
\begin{align}
\partial_{t}\gK_{ij} & =\Lie_{\gShiftVec}\gK_{ij}\signK\gD_{i}\gD_{j}\gLapse\isignK\gLapse\,\big[\gR_{ij}-2\gK_{ik}\gK^{k}{}_{j}+\gK\gK_{ij}\big]\nonumber \\
 & \qquad\signK\gLapse\,\gKappa\Big\{\,J_{ij}-\frac{1}{d-2}\Big(J-\rho\Big)\gSp_{ij}\,\Big\}\isignK\frac{1}{d-2}\gLapse\gSp_{ij}\gCC.\label{eq:GR-evol-ADM-K}
\end{align}
Hence, the ADM and the York equations only differ by an additive term
proportional to the scalar constraint $\gCC$, which vanishes for
a physical solution.\footnote{Note that $\gCC$ is usually removed during simplification (assuming
$\gCC=0$), which is forbidden here since $\gCC$ rather denotes the
\emph{violation} of the scalar constraint.} In other words, both systems are equivalent only in a subset of the
full space of solutions, called the constraint hypersurface (which
is a hypersurface in the space of solutions to the evolution equations).
As a consequence, although the two versions are physically equivalent,
they are not equivalent in their mathematical properties as shown
by Frittelli \cite{Frittelli:1996nj}. In particular, the constraint
evolution equations are well-posed for the York version (\ref{eq:g-evol})
and not well-posed for the ADM version (\ref{eq:GR-evol-ADM-K}).
The reason for a different dynamical behavior is that $\gCC$ contains
hidden second derivatives of the spatial metric (inside $\gR$) which
alter the hyperbolic structure of the differential equations. Furthermore,
Anderson and York \cite{Anderson:1998we} have shown that, assuming
a small (but nontrivial) change in the ADM action principle where
the independent gauge function is not taken to be the lapse $\gLapse$
but rather the densitized lapse $\gSector{\bar{\gLapse}}=\gLapse/\sqrt{\gSp}$
(the weight-minus-one lapse, also called the ``slicing density''),
the resulting evolution equations for the new momenta $\gSector{\bar{\pi}}^{ij}$
correspond precisely to those of York.\footnote{As Alcubierre \cite{Alcubierre:2012intro} noted, perhaps the densitized
lapse is more fundamental than the lapse itself.} 

We now turn to the HR theory and prove the following statement.

\begin{proposition}\label{prop:causal-cc}If the evolution equations
and the conservation laws hold, then the constraints (\ref{eq:g-cc})
satisfy a  homogeneous first-order symmetric hyperbolic system. If
initially satisfied, then the constraints hold at all times. The characteristic
cone of the system comprises the null cones of the two metrics in
the tangent space.\end{proposition}

\noindent \emph{Proof}. Assuming that the evolution equations $\tud{\gCE}ij=0$
and $\tud{\fCE}ij=0$ hold, the evolution equations for the constraint
violations reads,\bSe\label{eq:bim-CC-evol}
\begin{align}
\partial_{t}\gCC & =\Lie_{\gShiftVec}\gCC-\gLapse\gD_{i}\gCC^{i}-2\gCC^{i}\gD_{i}\gLapse\isignK\gLapse\gK\gCC,\\
\partial_{t}\gCC^{i} & =\Lie_{\gShiftVec}\gCC^{i}-\gCC\gD_{i}\gLapse-\gLapse\gD^{i}\gCC\isignK2\gLapse\tud{\gK}ij\gCC^{j}\isignK\gLapse\gK\gCC^{i},\\
\partial_{t}\fCC & =\Lie_{\fShiftVec}\fCC-\fLapse\fD_{i}\fCC^{i}-2\fCC^{i}\fD_{i}\fLapse\isignK\fLapse\fK\fCC,\\
\partial_{t}\fCC^{i} & =\Lie_{\fShiftVec}\fCC^{i}-\fCC\fD_{i}\fLapse-\fLapse\fD^{i}\fCC\isignK2\fLapse\tud{\fK}ij\fCC^{j}\isignK\fLapse\fK\fCC^{i}.
\end{align}
\eSe This is a system of $2d$ first-order partial differential equations
(PDE) in $2d$ unknown scalar functions $(\gCC,\gCC^{i},\fCC,\fCC^{i})$.
The system is obtained similarly to (\ref{eq:CC-evol}), where we
raised the indices of the vector constraints $\gCC^{i}=\gSp^{ij}\gCC_{j}$
and $\fCC^{i}=\fSp^{ij}\fCC_{j}$ for convenience. Neglecting the
homogeneous terms, the principal part of (\ref{eq:bim-CC-evol}) has
the following form,
\begin{equation}
(A^{\mu}\tud )IJ\partial_{\mu}u^{J}=\begin{pmatrix}\partial_{t}\gCC-\gShift^{k}\partial_{k}\gCC+\gLapse\partial_{k}\gCC^{k}\\
\partial_{t}\gCC^{i}-\gShift^{k}\partial_{k}\gCC^{i}+\gLapse\gSp^{ik}\partial_{k}\gCC\\
\partial_{t}\fCC-\fShift^{k}\partial_{k}\fCC+\fLapse\partial_{k}\fCC^{k}\\
\partial_{t}\fCC^{i}-\fShift^{k}\partial_{k}\fCC^{i}+\fLapse\fSp^{ik}\partial_{k}\fCC
\end{pmatrix}\!,\quad u^{I}\coloneqq\begin{pmatrix}\gCC\\
\gCC^{i}\\
\fCC\\
\fCC^{i}
\end{pmatrix}\!,\label{eq:cc-princ-part}
\end{equation}
where $I,J=1,...,2d$. The matrices in $\boldsymbol{A}^{\mu}\partial_{\mu}\boldsymbol{u}=\boldsymbol{A}^{0}\partial_{t}\boldsymbol{u}+\boldsymbol{A}^{k}\partial_{k}\boldsymbol{u}$
are easily identified as,
\begin{equation}
(A^{0}\tud )IJ\equiv\delta_{J}^{I},\quad\text{and}\quad(A^{k}\tud )IJ=\begin{pmatrix}-\gShift^{k} & \gLapse\delta_{j}^{k}\\
\gLapse\gSp^{ik} & -\gShift^{k}\delta_{j}^{i}
\end{pmatrix}\oplus\begin{pmatrix}-\fShift^{k} & +\fLapse\delta_{j}^{k}\\
\fLapse\fSp^{ik} & -\fShift^{k}\delta_{j}^{i}
\end{pmatrix}\!.\label{eq:cc-matrix-A}
\end{equation}
The system is clearly symmetric hyperbolic and well-posed. Therefore,
the stable propagation of the constraints holds. This concludes the
first part of the proof.

The principal symbol of (\ref{eq:bim-CC-evol}) is defined as the
matrix,
\begin{equation}
\boldsymbol{A}(\qvf)\coloneqq\boldsymbol{A}^{\mu}\qvf_{\mu}=\begin{pmatrix}\qvf_{0}-\gShift^{k}\qvf_{k} & \gLapse\qvf_{j}\\
\gLapse\gSp^{ik}\qvf_{k} & (\qvf_{0}-\gShift^{k}\qvf_{k})\delta_{j}^{i}
\end{pmatrix}\oplus\begin{pmatrix}\qvf_{0}-\fShift^{k}\qvf_{k} & \fLapse\qvf_{j}\\
\fLapse\fSp^{ik}\qvf_{k} & (\qvf_{0}-\fShift^{k}\qvf_{k})\delta_{j}^{i}
\end{pmatrix}\!,\label{eq:cc-princ-symb}
\end{equation}
where $\qvf_{\mu}\coloneqq(\qvf_{0},\qvf_{i})$ is an arbitrary covector.
 The characteristic polynomial of (\ref{eq:bim-CC-evol}) is defined
as the determinant of the principal symbol,  $P(\qvf)\coloneqq\det\boldsymbol{A}(\qvf)$,
evaluated as,\footnote{$\ \det\begin{psmallmatrix}A & B\\C & D\end{psmallmatrix}=\det(A-BD^{-1}C)\det D$.}
\begin{align}
P(\qvf) & =(\qvf_{0}-\gShift^{k}\qvf_{k})^{d-2}\Big[(\qvf_{0}-\gShift^{k}\qvf_{k})^{2}-\gLapse^{2}\qvf_{i}\gSp^{ik}\qvf_{k}\Big]\nonumber \\
 & \qquad\times(\qvf_{0}-\fShift^{k}\qvf_{k})^{d-2}\Big[(\qvf_{0}-\fShift^{k}\qvf_{k})^{2}-\fLapse^{2}\qvf_{i}\fSp^{ik}\qvf_{k}\Big].
\end{align}
Observe that the factor in the $\gMet$-sector can be rewritten by
noting that,
\begin{equation}
\gMet^{\mu\nu}\qvf_{\mu}\qvf_{\nu}=-\gLapse^{-2}\Big[(\qvf_{0}-\gShift^{k}\qvf_{k})^{2}-\gLapse^{2}\qvf_{i}\gSp^{ik}\qvf_{k}\Big].
\end{equation}
Thus, the characteristic polynomial has the form,
\begin{equation}
P(\qvf)=\gLapse^{2}\fLapse^{2}\,(\qvf_{0}-\gShift^{k}\qvf_{k})^{d-2}(\qvf_{0}-\fShift^{k}\qvf_{k})^{d-2}(\gMet^{\mu\nu}\qvf_{\mu}\qvf_{\nu})(\fMet^{\rho\sigma}\qvf_{\rho}\qvf_{\sigma}).\label{eq:princ-poly}
\end{equation}
The characteristics are obtained from $P(\qvf)=0$, where the roots
$\qvf_{0}$ are interpreted as the characteristic speeds \cite[p.\,582]{Courant:1962vol2}.
The characteristics of the propagation of the constraints are surfaces
with normal covectors $\qvf_{\mu}$ that satisfy either (i) $\qvf_{0}=\gShift^{k}\qvf_{k}$,
(ii) $\qvf_{0}=\fShift^{k}\qvf_{k}$, (iii) $\gMet^{\mu\nu}\qvf_{\mu}\qvf_{\nu}=0$,
or (iv) $\fMet^{\rho\sigma}\qvf_{\rho}\qvf_{\sigma}=0$. The characteristics
(i) and (ii) are timelike with respect to $\gMet$ and $\fMet$, and
tangent to the respective timelike unit normals $n_{\gMet}$ and $n_{\fMet}$
of the metrics. The characteristics (iii) and (iv) are null and propagate
on the null cones of $\gMet$ and $\fMet$ in the tangent space. Note
that the system (\ref{eq:bim-CC-evol}) is not strictly hyperbolic
since there are multiple roots in (i) and (ii). Nevertheless, the
system is strongly hyperbolic as one can determine a complete set
of eigenvectors for the principal symbol.  \hfill$\square$\bigskip{}

\noindent As shown, the causal propagation of the constraints in
the HR theory works almost the same as the one in GR \cite{Frittelli:1996nj},
provided that one crucial condition is fulfilled: The spatial metrics
$\gSp$ and $\fSp$ must be simultaneously positive definite. Such
a condition is guaranteed using the parametrization from \cite{Kocic:2018ddp},
supported by the theorem from \cite{Hassan:2017ugh}. 

If we started from the ADM version (\ref{eq:GR-evol-ADM-K}), the
terms $\gLapse\gD^{i}\gCC$ and $\fLapse\fD^{i}\fCC$ would be absent
from (\ref{eq:bim-CC-evol}), as well as the elements $\gLapse\gSp^{ik}\qvf_{k}$
and $\fLapse\fSp^{ik}\qvf_{k}$ from (\ref{eq:cc-princ-symb}). Then,
the principal symbol $\boldsymbol{A}(\qvf)$ would have two Jordan
blocks of size two, which implies that the system is weakly hyperbolic
since the principal symbol does not have a complete set of eigenvectors.
Consequently, the Cauchy problem is not well-posed for the ADM version,
and the free evolution (where the constraints are solved only to get
the initial data) is not guaranteed to give a solution to the HR bimetric
field equations (similarly to GR \cite{Frittelli:1996nj,Baumgarte:2010numerical}).

\begin{figure}
\noindent \centering{}\hspace{3mm}\begin{tikzpicture}[x=1mm,y=1mm]
  \node [anchor=south] at (0,-5) {\includegraphics[width=140mm]{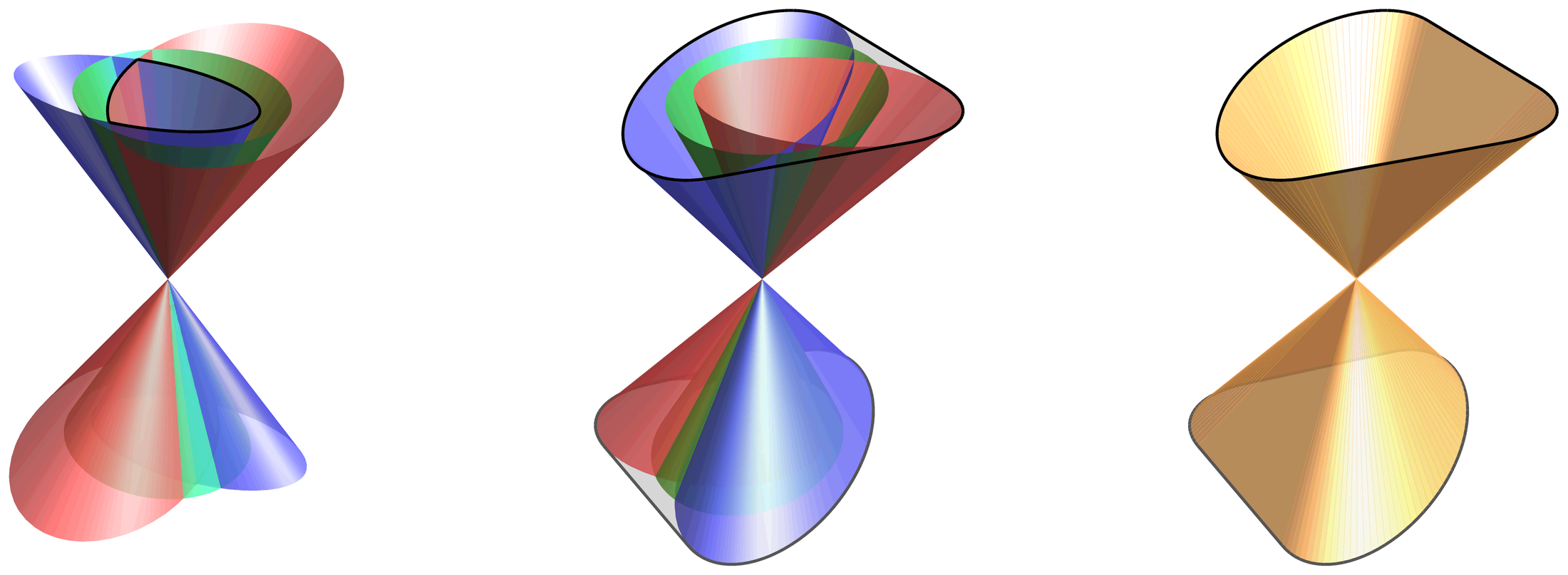}};
  \node at (-72,0) {(a)};
  \node at (-15,0) {(b)};
  \node at (38,0) {(c)};
\end{tikzpicture}
\vspace{-2ex}\caption{\label{fig:ccones}The normal cone (a) of the evolution equations
for the constraint violations in the cotangent space, and the corresponding
ray cone (b) in the tangent space. The geometric mean cones (shown
in green) are not part of the normal or the ray cone.  The causal
cone (c) is the convex hull of the causal cones of of the two individual
systems, which encloses the ray cones (b).}
\end{figure}
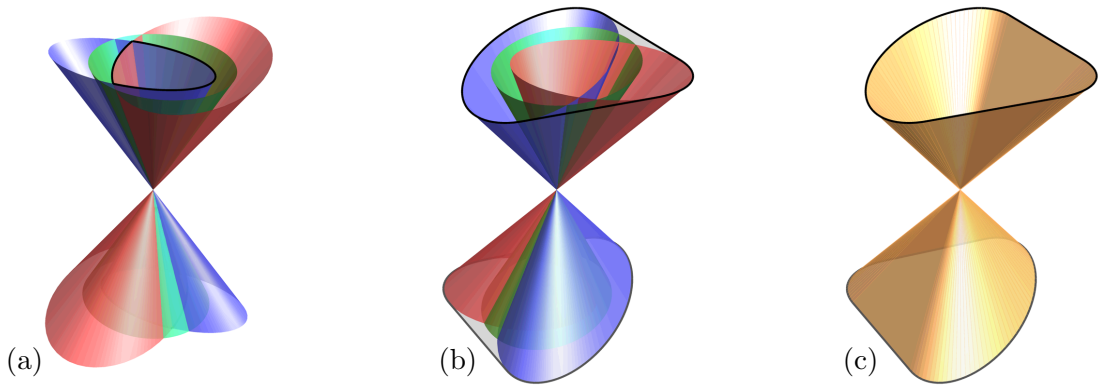

We now illustrate the characteristics of the PDE system (\ref{eq:bim-CC-evol}).
The attributes \emph{timelike}, \emph{spacelike}, \emph{null}, or
\emph{causal} will be used to state the causal structure defined by
the PDE. The causality relative to metrics will be stated explicitly.
For comparison, the causal structure of several physical systems with
nontrivial principal polynomials is given in appendix \ref{app:hp-examples}.
The \emph{normal cone} of the system is defined by the characteristic
polynomial (\ref{eq:princ-poly}) in the cotangent space as shown
in Figure~\ref{fig:ccones}a (where the planes $\qvf_{0}=\gShift^{k}\qvf_{k}$
and $\qvf_{0}=\fShift^{k}\qvf_{k}$ are suppressed). 
 The \emph{ray cone} is defined by the characteristic vectors in the
tangent space, shown in Figure~\ref{fig:ccones}b. In the case of
(\ref{eq:princ-poly}), the \emph{causal cone} of the combined system
is the convex hull of the causal cones of the two subsystems, that
is, the set of all sums of the form $\xi_{\gMet}^{\mu}+\xi_{\fMet}^{\mu}$,
with $\xi_{\gMet}^{\mu}$ in the null cone of $\gMet_{\mu\nu}$ and
$\xi_{\fMet}^{\mu}$ in the null cone of $\fMet_{\mu\nu}$ \cite{Geroch:2010da,Lax:2006hPDE,Rauch:2012hyp}.
The convex inner mantle of the normal cone is associated with the
convex hull of the ray surface (both are indicated by thicker lines
in Figure~\ref{fig:ccones}). The outer shell of the ray cone is
not necessarily convex and therefore it does not coincide with its
convex hull.  The signal propagation for the combined system is in
those spacetime directions obtained by taking sums of the signal-propagation
directions for the two systems separately. The outer sheet of the
causal cone defines the domain of dependence (called the domain of
determinacy of the initial manifold in \cite[p.\,439]{Courant:1962vol2}),
where upon the dynamical evolution of the initial data, the constraints
are satisfied at all points inside the domain.  If the two individual
subsystems (sectors) had not been interacting, $V(S)=0$, there would
be no need for concerning the convex hull. Since the two sectors are
coupled (the lapses and shifts are not are not independent of each
other), then the ray surface remains unchanged, but the data must
be given in the convex hull (see the footnote on p.\,652 in \cite{Courant:1962vol2}).
Note that each FOSH system carries within itself its own initial-value
formulation with its own causal cones for signal propagation (rooted
in the PDE structure itself). By combining several FOSH systems (turning
on interactions between them but not corrupting the highest-order
derivatives), the causal cones are also combined. Importantly, all
the combined systems appear on an equal footing. This formulation
manifests what Geroch \cite{Geroch:2010da} calls the \emph{democracy
of causal cones}: No subsystem (or a set of causal cones) has priority
over any others.

\section{Summary and outlook}

\label{sec:discussion}

The goal was to investigate the causal properties of the constraint
evolution equations in the ghost-free bimetric theory. This resulted
in a proposition that, if the evolution equations are given in standard
3+1 form, the propagation of the constraints satisfies a first-order
symmetric hyperbolic system. To ensure the well-posedness of the constraint
evolution equations and the stable propagation of the constraints
is necessary, otherwise the unphysical modes will not be bounded but
amplified during the free evolution. Some other aspects of the causality
in bimetric theory have been investigated in \cite{Izumi:2013poa,Camanho:2016opx,Bellazzini:2017fep,Hinterbichler:2017qcl,Bonifacio:2017nnt,Hinterbichler:2017qyt}.

The causal structure appearing in the proof of Proposition~\ref{prop:causal-cc}
can be related with the analysis of Schuller et al\@.~\cite{Schuller:2016onj,Schuller:2014jia},
who studied gravitational dynamics and partial differential equations
for arbitrary tensorial spacetimes carrying predictive, interpretable,
and quantizable matter. These three requirements can be translated
into the corresponding algebraic conditions on the underlying geometry,
which state that the geometry must be bi-hyperbolic, time-orientable,
and energy-distinguishing. The authors of \cite{Schuller:2016onj}
further investigated the gravitational closure of two Klein\textendash Gordon
scalar fields on a bimetric geometry, pointing out that the principal
polynomial of such a theory is the product of the principal polynomials
of the two individual scalar fields {[}ibid.~appendix~B{]}. This
is the same algebraic structure as found in the principal polynomial
(\ref{eq:princ-poly}) of the evolution equations for the constraint
violations (\ref{eq:bim-CC-evol}). Moreover, the bi-hyperbolicity
condition is equivalent to the requirement that the null cones of
the metrics $g$ and $f$ intersect in such a way that provides the
existence of the real principal square root of $g^{-1}f$ \cite{Hassan:2017ugh}.
Therefore, the HR bimetric spacetime can carry predictive, interpretable,
and quantizable matter.

Note that the HR evolution equations (\ref{eq:g-evol}) are not themselves
a FOSH system. Hence, it would be desirable to cast the HR \nPlusOne{}
system into a symmetric hyperbolic form ensuring the well-posedness
of the full theory. As a guideline, we already encountered one observation
that comes out from the proof of Proposition~\ref{prop:causal-cc}.
The causal propagation of the constraints in the HR theory works almost
the same as the one in GR due to the theorem from \cite{Hassan:2017ugh}
and the property that the evolution equations of the two sectors are
coupled algebraically. This means that some of the \nPlusOne{} hyperbolic
reductions in GR are  applicable to the HR theory, such as the Frittelli\textendash Reula
\cite{Frittelli:1996wr} formulation, or the Einstein\textendash Christoffel
formulation of Anderson and York \cite{Anderson:1999qx} which employs
the earlier mentioned densitized lapse function.\footnote{In fact, another more recent approach by Kidder et al.~\cite{Kidder:2001tz}
shows that densitization of the lapse is a necessary condition for
the hyperbolicity. The Kidder\textendash Scheel\textendash Teukolsky
formulation \cite{Kidder:2001tz} comprises both Anderson\textendash York
\cite{Anderson:1999qx} and Frittelli\textendash Reula \cite{Frittelli:1996wr}
formulations as subsets. }

As a motivation, let us address the question about the well-posedness
of the reduced Einstein\textendash matter equations, discussed in
\cite{Rendall:2008pdegr} and \cite{Friedrich:2000qv}. Consider
the reduced Einstein equations written as a FOSH system with basic
unknowns denoted by $u$. Suppose that the matter equations can be
written in a symmetric hyperbolic form with basic unknowns $v$. If
the coupling of these systems is only by terms of order zero, then
the combined system is symmetric hyperbolic for $(u,v)$. The necessary
condition for this is that the matter equations contain at most first
derivatives of the metric (e.g., the Christoffel symbols) and that
the stress-energy tensor contains no derivatives of $u$. In the bimetric
case, the `matter' is another Einstein sector already in a FOSH form,
therefore we would have the reduced Einstein\textendash Einstein system.
Since the bimetric stress-energy tensors are purely algebraic with
respect to the metrics, the above conditions for the Einstein\textendash matter
system can be satisfied, so a local existence theorem for the reduced
Einstein\textendash Einstein bimetric system is obtainable. However,
the importance of such results should not be overestimated when considering
numerical applications since the hyperbolization of the evolution
equations does not necessarily contribute to numerical accuracy and
stability.\footnote{See \cite{Shinkai:2002yf} for a review of the efforts to reformulate
the Einstein equations for stable numerical simulations.} Adapting, for instance, the BSSN formulation \cite{Shibata:1995we,Baumgarte:1998te}
might be more attractive for numerical bimetric relativity.\footnote{Such a modification could also be implemented as a component of the
Einstein Toolkit \cite{Loffler:2011ay} which is based on the the
BSSN evolution system.}

\acknowledgments\vspace{-1.8ex}

I am grateful to Fawad Hassan and Francesco Torsello for numerous
fruitful discussions and reading of the manuscript. 

\clearpage


\emitAppendix

\section{Hyperbolic polynomials}

\label{app:hp-examples}

\begin{figure}[H]
\noindent \centering{}\begin{tikzpicture}[x=1mm,y=1mm]
  \node [anchor=south] at (0,-5) {\includegraphics[width=147mm]{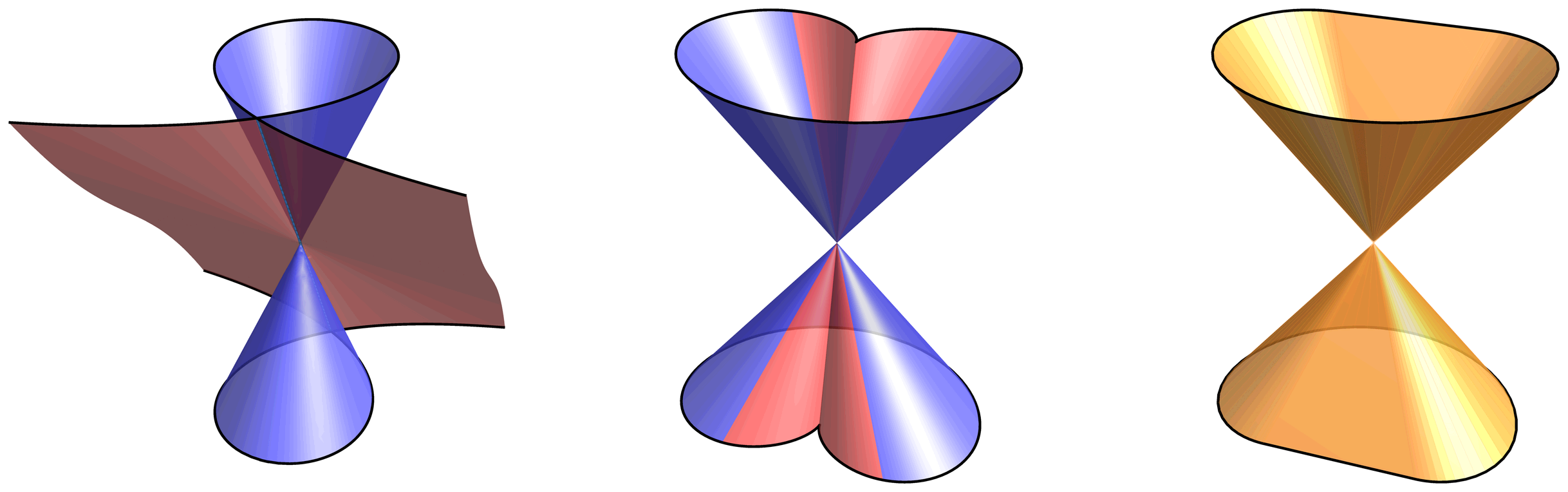}};
  \node at (-57,0) {(a)};
  \node at (-14,0) {(b)};
  \node at (36,0) {(c)};
\end{tikzpicture}
\vspace{-2ex}\caption{\label{fig:ccones3} The normal cone (a), the ray cone (b), and the
causal cone (c) for a peculiar hyperbolic PDE of third order. The
reciprocal sheets of the normal and the ray cone are depicted in the
same color. See appendix \ref{app:hp-examples} for further explanations.}
\end{figure}
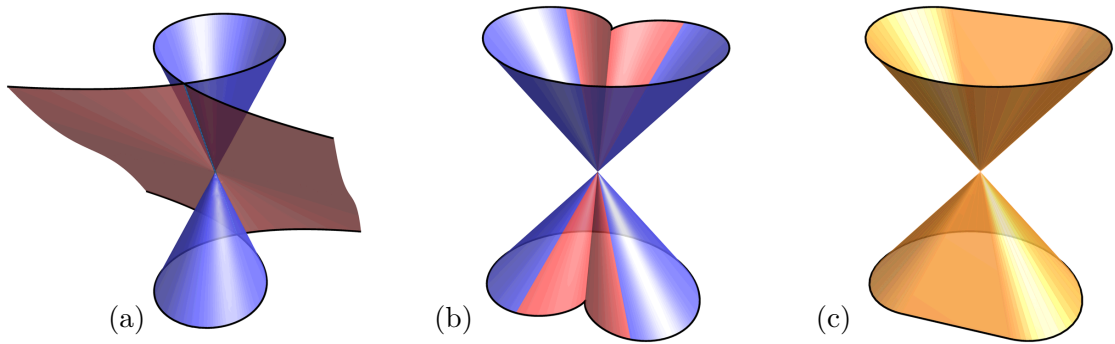

In this section, we illustrate possible peculiarities in the geometrical
structure of hyperbolic polynomials. We first consider a somewhat
artificial example studied on page 588 in \cite{Courant:1962vol2}.
The PDE of interest is of third order,
\begin{equation}
\left[\left(\partial_{t}-\partial_{y}\right)\left(\partial_{t}+\partial_{y}\right)^{2}-\partial_{x}^{2}\left(2\partial_{t}+\partial_{y}\right)\right]u(t,x,y)=0,
\end{equation}
with the principal polynomial $\left(\xi_{0}-\xi_{2}\right)\left(\xi_{0}+\xi_{2}\right)^{2}-\xi_{1}^{2}\left(2\xi_{0}+\xi_{2}\right)$
which can be written,
\begin{equation}
P(\xi)=\left(\xi_{0}^{2}-\xi_{1}^{2}-\xi_{2}^{2}\right)\left(\xi_{0}+\xi_{2}\right)-\xi_{1}^{2}\xi_{0}.\label{eq:ap-P1}
\end{equation}
The normal cone is defined by $P(\xi)=0$ in the $(\xi_{0},\xi_{1},\xi_{2})$
space, shown in Figure~\ref{fig:ccones3}a. Note that, if the term
$\xi_{1}^{2}\xi_{0}$ is absent from (\ref{eq:ap-P1}), the normal
cone degenerates into the plane $\xi_{0}+\xi_{2}=0$ tangent to the
null cone $\xi_{0}^{2}-\xi_{1}^{2}-\xi_{2}^{2}=0$. The normal surface
is obtained at $\xi_{0}=-1$ in the $(\xi_{1},\xi_{2})$-plane solving
$P(-1,\xi_{1},\xi_{2})=0$. This is the curve of third order (the
folium of Descartes) indicated by thicker lines in Figure~\ref{fig:ccones3}a.
The ray cone is obtained by solving $\dot{x}^{\mu}=\partial P/\partial\xi_{\mu}$
where $\xi$ satisfies $P(\xi)=0$. The ray cone and ray surface are
shown in Figure~\ref{fig:ccones3}b. The causal cone is the convex
hull of the ray cone, shown in Figure~\ref{fig:ccones3}c. 

\begin{figure}
\noindent \centering{}\begin{tikzpicture}[x=1mm,y=1mm]
  \node [anchor=south] at (0,-5) {\includegraphics[width=130mm]{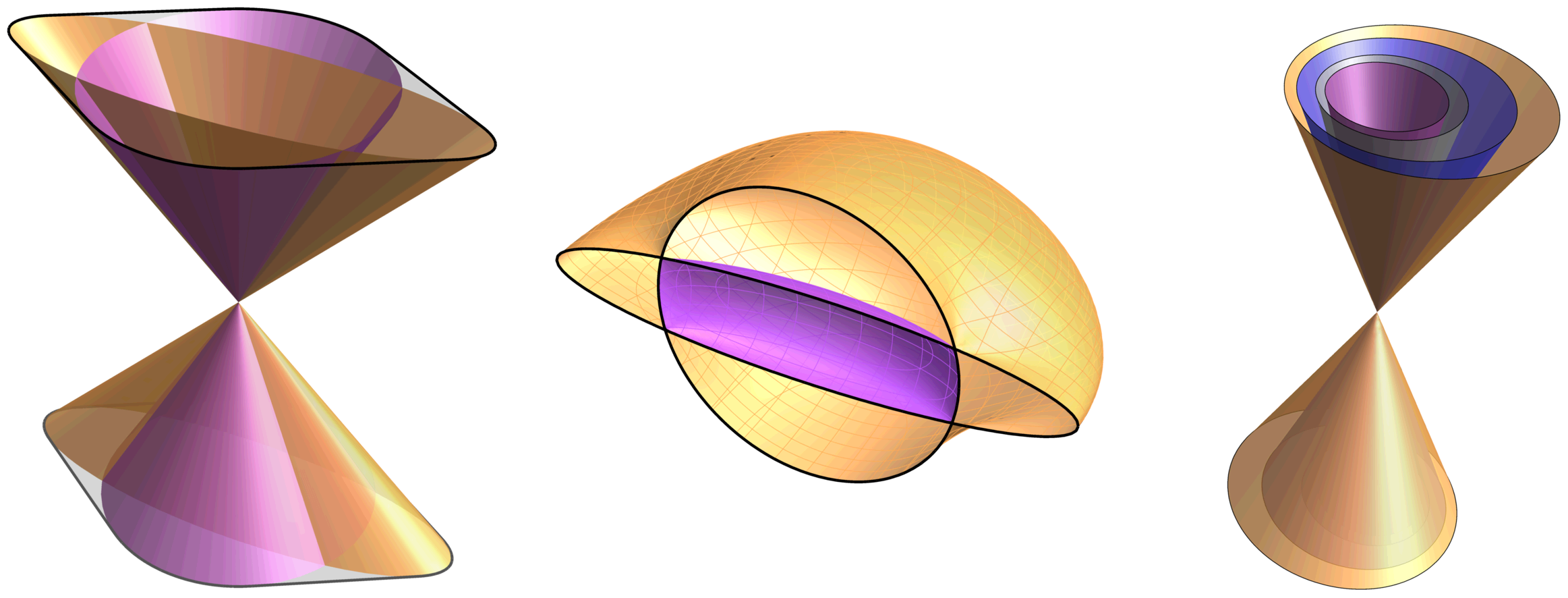}};
  \node at (-62,0) {(a)};
  \node at (-5,0) {(b)};
  \node at (32,0) {(c)};
\end{tikzpicture}
\vspace{-2ex}\caption{\label{fig:ccones2}Some hyperbolic systems of physical significance:
(a) the ray cone of Maxwell's equations for crystal optics in 2+1
dimensions, (b) the section of the ray surface for the crystal optics
example in 3+1 dimensions, and (c) a symmetric hyperbolic system obtained
from the 3+1 form of the Einstein field equations. See appendix \ref{app:hp-examples}
for references.}
\end{figure}
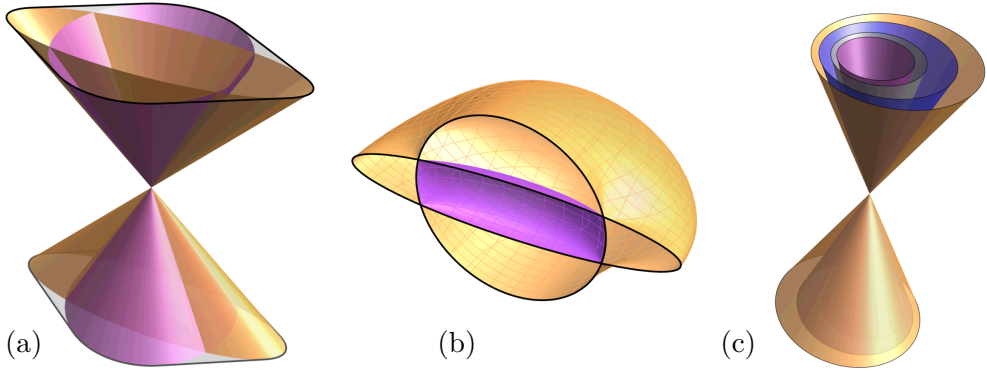

\noindent This example reveals two possible peculiar features of
hyperbolic polynomials: 
\begin{enumerate}[label=(\roman*),topsep=2pt,partopsep=0pt,itemsep=0pt,parsep=0pt]
\item The normal surface can have double points, in which case a lid must
be added to form the convex hull as the ray cone may not be convex.
\item The normal surface may have a sheet extending to infinity (or a point
of inflection). In such a case, the ray surface will have a cusp at
an isolated point. 
\end{enumerate}
Furthermore, chapter VI in \cite{Courant:1962vol2} investigates several
systems of physical significance which posses the above properties,
for instance, Maxwell's equations in crystal optics which exhibit
a feature of an anisotropy where the velocity of a propagating wave
depends upon the direction of propagation. In that example, the normal
surface and the ray surface are surfaces of the fourth-degree. The
ray cone for the crystal optics PDE reduced to 2+1 dimensions is shown
in Figure~\ref{fig:ccones2}a. The section of the ray surface for
the full 3+1-dimensional case of the reduced Maxwell\textquoteright s
equations is shown in Figure~\ref{fig:ccones2}b.

Finally, we show the causal structure of a symmetric hyperbolic system
obtained from the 3+1 equations in GR \cite{Friedrich:2000qv}. The
characteristics of that system propagate on several null cones, shown
in Figure~\ref{fig:ccones2}c. In this example, the null cone of
the spacetime metric is not necessarily the outer one, that is, the
causal cone is wider than the null cone of the metric.

\section{\nPlusOne{} bimetric variables}

\label{app:bim-vars}

The primary variables in the parametrization \cite{Kocic:2018ddp}
are the lapses $\gLapse$, $\fLapse$, the spatial vielbeins $\tud{\gE}ai$,
$\tud{\fE}ai$, the overall shift vector $\hShiftVec^{i}$ of the
geometric mean, and the Lorentz vector $\sLp^{a}$ that defines the
separation between the two metrics. The separation $\sLp$ is in fact
the boost parameter of the Lorentz transformation that comprises the
spatial part $\sLs$ and the Lorentz factor $\sLt$ where $\sLp=\sLs\sLv=\sLt\sLv$
and,
\begin{align}
\sLt & \coloneqq\big(1+\sLp^{{\scriptscriptstyle \tr}}\sEta\sLp\big)^{1/2}=\big(1-\sLv^{{\scriptscriptstyle \tr}}\sEta\sLv\big)^{-1/2},\\
\sLs & \coloneqq\big(\sI+\sLp\sLp^{\tr}\sEta\big)^{1/2}=\big(\sI-\sLv\sLv^{\tr}\sEta\big)^{-1/2}.
\end{align}
The rest of the variables (\ref{eq:Np1-vars}) are derived from $\gLapse$,
$\tud{\gE}ai$, $\fLapse$, $\tud{\fE}ai$, $\sLp^{a}$ , and $\hShiftVec^{i}$
as,
\begin{align}
\gSp & \coloneqq\gE^{{\scriptscriptstyle \tr}}\sEta\gE, & \fSp & \coloneqq\fE^{{\scriptscriptstyle \tr}}\sEta\fE,\\
\sgn & \coloneqq\gE^{-1}\sLv, & \sfn & \coloneqq\fE^{-1}\sLv,\\
\gShiftVec & \coloneqq\hShiftVec+\gLapse\sgn, & \fShiftVec & \coloneqq\hShiftVec-\fLapse\sfn,\\
\sgQ & \coloneqq\gE^{-1}\sLs^{2}\gE, & \sfQ & \coloneqq\fE^{-1}\sLs^{2}\fE,\\
\sgD & \coloneqq\fE^{-1}\sLs^{-1}\gE, & \sfD & \coloneqq\gE^{-1}\sLs^{-1}\fE,\\
\sgB & \coloneqq\sgD^{-1}=\gE^{-1}\sLs\fE=\sfD\sfQ=\sgQ, & \sfB & \coloneqq\sfD^{-1}=\fE^{-1}\sLs\gE=\sgD\sgQ=\sfQ,\\
\sgV & \coloneqq\usignV\betaSum\,e_{n}(\sfD), & \sfV & \coloneqq\usignV\betaSum\,\sLtinv e_{n-1}(\sgB),\\
\sgU & \coloneqq\usignV\betaSum\,\sLtinv Y_{n-1}(\sgB), & \sfU & \coloneqq\usignV\betaSum\,\sfD\,Y_{n-1}(\sfD),\\
\sgQU & \coloneqq\usignV\betaSum\,\sgB\,Y_{n-1}(\sfD), & \sfQU & \coloneqq\usignV\betaSum\,\sLtinv\sfQ\,Y_{n-1}(\sgB).
\end{align}


\clearpage
\ifprstyle
  \bibliographystyle{apsrev4-1}
  \bibliography{prop-constr}
\else
  \bibliographystyle{JHEP}
  \bibliography{prop-constr}
\fi

\end{document}